\documentclass[useAMS]{mn2e}
\usepackage{epsfig}

\newcommand{\Msol}{M$_{\odot}$}
\newcommand{\HI}{H\,{\sc {i}}~}

\newcommand{\Msold}{M$_{\odot}$\,yr$^{-1}$}

\newcommand{\Vexp}{V$_{\rm exp}$}

\newcommand{\Tstar}{T$_{\rm \star}$}
\newcommand{\kms}{km\,s$^{-1}$}

\newcommand{\lsim}{~\rlap{$<$}{\lower 1.0ex\hbox{$\sim$}}}
\newcommand{\gsim}{~\rlap{$>$}{\lower 1.0ex\hbox{$\sim$}}}

\title[\HI 21-cm line profile]{Modeling the \HI 21-cm line profile from 
circumstellar shells around red giants}
\author[D. T. Hoai et al.]
{D. T. Hoai$^{1,2}$, P. T. Nhung$^{1,2}$, E. G\'erard$^{3}$, L. D. 
Matthews$^{4}$, E. Villaver$^{5}$,
\newauthor and T. Le Bertre$^{1}$\\
$^{1}$LERMA, UMR\,8112, CNRS \& Observatoire de Paris/PSL, 61 av. de 
           l'Observatoire, F-75014 Paris, France\\
$^{2}$VNSC/VAST, 18 Hoang Quoc Viet, Cau Giay, Hanoi, Vietnam\\
$^{3}$GEPI, UMR\,8111, CNRS \& Observatoire de Paris/PSL, 5 place J. Janssen, 
           F-92195 Meudon Cedex, France\\
$^{4}$MIT Haystack Observatory, Off Route 40, Westford, MA 01886, USA\\
$^{5}$Departamento de F\'isica Te\'orica, Universidad Aut\'onoma de Madrid, 
Cantoblanco 28049 Madrid, Spain\\}
\begin{document}

\date{Accepted, 3 March 2015. Received, 3 March 2015; in original form, 
2 February 2015}
%\date{\today}
\pagerange{\pageref{firstpage}--\pageref{lastpage}} \pubyear{2002}
\maketitle
\label{firstpage}

\begin{abstract}
We present \HI line profiles for various models of circumstellar shells 
around red giants. In the calculations we take into account the effect of the 
background at 21 cm, and show that in some circumstances it may have an 
important effect on the shape and intensity of the observed line profiles. 
We show that self-absorption should also be considered depending on 
the mass loss rate and the temperature reached by circumstellar gas.

\HI emission from circumstellar shells has been mostly reported from 
stars with mass loss rates around 10$^{-7}$ \Msold. We discuss the possible 
reasons for the non detection of many sources with larger mass loss rates 
that are hallmarks of the end of the AGB phase. Although radiative 
transfer effects may weaken the line emission, they cannot alone account 
for this effect. Therefore, it seems likely that molecular hydrogen, 
rather than atomic hydrogen, dominates the composition of matter expelled 
by stars at the end of their evolution on the Asymptotic Giant Branch. 
However sensitive \HI observations can still yield important 
information on the kinematics and physical properties of the circumstellar 
material at large distances from central stars with heavy mass loss, 
despite the low abundance of atomic hydrogen.
\end{abstract}


\section{Introduction}
Asymptotic giant branch (AGB) stars are undergoing mass loss via slow  
stellar winds (Olofsson 1999). They get surrounded by expanding circumstellar 
shells that are well characterized by CO emission lines in the 
millimeter range (Knapp \& Morris 1983, Loup et al. 1993). 
From the modeling of the CO line profiles, expansion velocities are 
derived to be  in the range from a few to 40 \kms, and mass loss rates, 
from $\sim$ 10$^{-7}$ to a few 10$^{-4}$ \Msold. 

Most of the matter in circumstellar shells should be in hydrogen ($\sim$70\,\% 
in mass). Early searches for 21-cm line emission from atomic hydrogen were 
mostly unsuccessful (Zuckerman et al. 1980, Knapp \& Bowers 1983) leading to 
the conclusion that hydrogen in stellar envelopes is molecular rather than 
atomic. However, more recently, the \HI line at 21 cm has been detected from 
several AGB stars (G\'erard \& Le\,Bertre 2006). A number of the detected 
stars have also been imaged in \HI with the Very Large Array (VLA, Matthews 
\& Reid 2007, Matthews et al. 2008, 2011, 2013). 
In general, the detected sources have mass loss rates $\sim\,10^{-7}$\,\Msold, 
and they show a narrow \HI line profile (3-5 \kms). On the other hand, many 
sources with large mass loss rate ($\ge\,10^{-6}$\,\Msold) remain undetected. 

Again, a possible explanation could be that hydrogen in sources with large 
mass loss rates is mostly in molecular form. 
However, this is not entirely satisfactory 
for at least two reasons. First, molecular hydrogen is expected to be 
photo-dissociated by UV photons from the interstellar radiation field 
(ISRF) in the external parts of circumstellar shells (Morris \& Jura 1983). 
Second, the mass loss rate from AGB stars is supposed to be highly variable 
(e.g. Vassiliadis \& Wood 1993), so that a given star should be surrounded 
by interacting shells, originating from different episodes of mass loss 
at different rates, from 10$^{-8}$\,\Msold ~upwards, and different velocities 
(Villaver et al. 2002), with probably  a mix of atomic and molecular hydrogen. 

Another possibility is that radiative transfer effects are weakening the 
21-cm line emission, thus rendering difficult the detection of H\,{\sc {i}}. 
The narrowness of the line profiles could increase the optical depth at 
particular frequencies, thus exacerbating self-absorption. The 
presence of an intense background at 21 cm may also play a role (Levinson 
\& Brown 1980). In this paper, we examine the formation of the \HI line 
in circumstellar shells around AGB stars, taking into account these effects.

\section[]{Context}
\label{astroback}

Hydrodynamical models of dust-forming long-period variable stars reproduce 
outflows with mass loss rate ranging from $\sim 10^{-8}$ 
to $10^{-4}$ \Msold, and expansion velocity ranging from a few to 40 \kms 
~(Winters et al. 2000). Close to the central star, the density is much larger 
than in the interstellar medium (ISM), and the wind expands freely 
(freely expanding wind). We hereafter refer to the case where the wind 
expands freely as {\bf scenario\,1}. 
However, at some point, the stellar wind is slowed down by the ambiant 
medium, and a structure develops in which both circumstellar and interstellar 
matter accumulates. Such structures were discovered
with {\it IRAS} owing to the presence of spatially resolved 60$\mu$m
continuum emission (Young et al. 1993) and are referred to as
`detached shells'{\footnote{This designation has also been used for  
shells resulting from an outburst of mass loss and observed in CO molecular 
lines (Olofsson et al. 2000). In this paper, we adopt to the definition 
of Young et al.}}.

Subsequently, the \HI line at 21 cm was detected in the direction of several 
resolved {\it IRAS} sources (Le~Bertre \& G\'erard 2004; G\'erard \& Le~Bertre 
2006). The \HI emission is more extended than the CO emission and generally 
corresponds with the far-infrared size. Also the main components  
of the \HI line are narrower than the CO line profiles, 
thus demonstrating that stellar outflows are slowed down by an external 
medium, and that the \HI line profiles can be used to probe the kinematics 
in the external regions of circumstellar shells. 
Imaging obtained with the VLA has revealed detached \HI shells with narrow 
line profiles, sometimes associated with extended tails (Matthews et al. 2013).

Libert et al. (2007) developed a model of a  detached shell (hereafter 
{\bf scenario\,2}) adapted to the \HI emission from the prototypical case 
of Y CVn. In this model, a double-shock structure develops, 
with a termination shock, facing the central star, where the supersonic 
stellar wind is abruptly slowed down, 
and a leading shock (i.e. outward facing shock) where interstellar matter 
is compressed and integrated in the detached shell. These two limits define 
the detached shell which is thus composed of circumstellar and 
interstellar matter separated by a contact discontinuity. The circumstellar 
matter is decelerated, by a factor $\sim 4$, when it crosses the termination 
shock, and heated. It then cools down, while the expansion velocity further 
decreases and the density increases until the contact discontinuity. 
The model assumes spherical symmetry and stationarity. Adopting an 
arbitrary temperature profile, the equation of motion is solved numerically 
between the two limits of the detached shell. The \HI line profile 
is then calculated assuming that the emission is optically thin. 

Libert et al. (2007) could reproduce the \HI line profile observed in Y CVn 
and generalize the model to other cases of sources with mass loss rates 
of about 10$^{-7}$\,\Msold ~(e.g. Libert et al. 2010). 
However, as discussed in Matthews et al. (2013), although this model is 
effective in reproducing spatially integrated line profiles, it faces some 
difficulties in reproducing the spatial distribution of the \HI emission. 
Namely, at small radii, it seems to predict too much flux in both the shell 
component and the freely expanding wind component of the spectral profile.
The mass loss variations with time and the interaction between consecutive 
shells may need to be taken into account (see Villaver et al. 2002). 
The more elaborate numerical modeling performed by Villaver et al.  
follows the mass loss and velocity 
variations of the stellar winds of thermally pulsing AGB stars, and describes 
self-consistently the formation and development of circumstellar shells 
embedded in the ISM (hereafter {\bf scenario\,3}). They show that large 
regions (up to 2.5\,pc) of neutral gas may be formed around AGB stars. 

In this work, we will simulate the \HI line profiles that would be 
expected for these three kinds of scenarios and examine how they compare 
with observations presently available. 

\section[]{General comments on the formation of the \HI line profile}
\label{equations}

The \HI line at 21 cm is produced by a transition between two levels of the 
ground state of hydrogen. The rest frequency, $\nu_0$, corresponds to 
1420.4 MHz, and the transition probability, $A_{21}$, 
to 2.87 10$^{-15}$ s$^{-1}$. As the transition probability between the two 
levels is very low, the natural width of the line is extremely small, and 
the frequency dependence of the absorption and emission coefficients 
are entirely defined by the atomic hydrogen velocity. This is determined 
by the bulk motion of the gas (including, if present, turbulence), and 
the hydrogen kinetic temperature ($T$). We assume local thermal equilibrium 
(LTE). Also, in general, as the frequencies are small, one may replace 
$exp (h \nu / kT) -1$, by $h \nu / kT$  ($h \nu / k \sim$ 0.07\,K). 

The intensity emitted, at a frequency $\nu$, along an element $ds$, 
$dI_{\nu}$, is given by :
\begin{equation}
\frac{dI_{\nu}}{ds} = \frac{h \nu}{4 \pi} n_2(\nu) A_{21} 
\end{equation}
where $n_2(\nu)$ is the number density of hydrogen atoms in the upper level 
with a velocity along the line of sight corresponding to $\nu$.

Similarly the absorption part is given by :
\begin{equation}
\frac{dI_{\nu}}{ds} = -\kappa(\nu) I_{\nu}, 
\end{equation} 
with the absorption coefficient, $\kappa(\nu)$, given by :
\begin{equation}
\kappa(\nu) = \frac{c^2 n_1(\nu) g_2}{8 \pi {\nu_0}^2 g_1} A_{21} \frac{h \nu}{k T}
\end{equation} 
(cf. equation 3.14 in Lequeux 2005). $g_2$ and $g_1$ are the statistical 
weights of the upper and lower levels, respectively, and 
$n_1(\nu)$ is the number density of hydrogen atoms in the lower level with 
a velocity along the line of sight corresponding to $\nu$.
It is interesting to note that, in general, the absorption coefficient varies 
as $1/T$. 

For the \HI line at 21 cm: $g_1 = 1$, $g_2 = 3$, and for a population 
in statistical equilibrium, $n_1 = 1/4\,n_H$, and $n_2 = 3/4\,n_H$, 
where $n_H$ is the total number density of hydrogen atoms in the ground state 
($n_H = n_1 + n_2$), and assuming $exp (h \nu / kT) = 1$ for the Boltzmann 
factor. 

In practice, spectra are represented as a function of the velocity, $V$. 
We may thus replace $n_1(\nu)$ by $1/4\,n_H (V)$ and $n_2(\nu)$ by 
$3/4\,n_H (V)$. With this convention, 
the absorption coefficient can be rewritten as: 
\begin{equation}
\kappa(V) = \frac{3 c^2 n_H (V)}{32 \pi \nu_0} A_{21} \frac{h}{k T}.
\end{equation} 

Finally the radiative transfer equation is written as :
\begin{equation}
\frac{dI(V)}{ds} = \frac{3 h \nu_0}{16 \pi} n_H (V) A_{21} - 
\frac{3 c^2 n_H (V)}{32 \pi \nu_0} A_{21} \frac{h}{k T} I(V). 
\end{equation}

At radio frequencies, it is usual to express the intensity in terms of 
the equivalent temperature of a blackbody that would give the same intensity 
in the same spectral domain. With this convention, the boundary condition 
can be defined through a background brightness temperature, $T_{BG}$ : 
\begin{equation}
I^+(V) = \frac{2 k \nu_0^2}{c^2} T_{BG}(V), 
\end{equation}
$I^+$ refering to the incoming intensity on the rear side of the shell. 
The background is the sum of the 3-K cosmic emission, the synchrotron 
emission from the Galaxy and the \HI emission from the ISM 
located beyond the circumstellar shell with respect to the observer.  
The first component is a continuum emission, which is smooth, angularly and 
spectrally. The second component is also smooth spectrally, but it presents a 
strong dependence with galactic latitude, and shows also some sub-structures. 
The sum of these two components has been mapped with a spatial resolution of 
0.6$^{\circ}$ by Reich (1982), Reich \& Reich (1986) and Reich et al. (2001). 
The third component (\HI emission from the ISM) 
shows both strong spatial and spectral dependences, 
which make it a serious source of confusion. It has been mapped with 
a spatial resolution of 0.6$^{\circ}$ and a spectral resolution of 1.3 \kms 
~by Kalberla et al. (2005; Leiden-Argentina-Bonn, LAB, survey). 
Surveys of selected regions of the sky, in particular along the Galactic 
Plane, have been obtained at a resolution down to 1$'$, and show spatial 
structures, like filaments or clouds, at all sizes (e.g. Stil et al. 2006). 

Away from the Galactic Plane, typical values range from $T_{BG} \sim$3-5 K, 
outside the range of interstellar \HI emission, to $T_{BG} \sim$10-20 K 
inside an interstellar \HI emission. Close to the Galactic Plane, 
the continuum may reach $T_{BG} \sim$10-20 K, and, including 
the interstellar \HI emission, the background may reach $T_{BG} \sim$100 K. 
Note that the background temperature, $T_{BG}$, is not directly related 
to the kinetic temperature of the surrounding ISM. 

In addition, in some cases, a radio-source may be seen in the direction of a 
circumstellar shell. In such a case, we have an unresolved continuum emission 
(see e.g. Matthews et al. 2008). Such a source may be useful to probe 
the physical conditions within the circumstellar shell in a pencil-beam mode. 

\section[]{Simulations}
\label{simul}

For the present work, we adapted the code developed by Hoai et al. (2014). 
It is a ray-tracing code that takes into account absorption and emission 
in the line profile. It can handle any kind of geometry, but for the  
purpose of this paper we restricted our simulations to circumstellar shells 
with a spherical geometry as described in Sect.\,\ref{astroback}. 
We assume that, in each cell, the gas is in equilibrium and that 
the distribution of the velocities is maxwellian. 

In this section, we explore the line profiles for a source that is not 
resolved spatially by the telescope, and assume a uniform response in 
the telescope beam (boxcar response, cf. Gardan et al. 2006). 
We also assume that the line profiles can be extracted from position-switched 
observations, i.e. that there is no spatial variation of the background. 
The flux densities are expressed in the
units of Janskys (Jy), where 1 Jy\,=\,10$^{-26}$\,W\,m$^{-2}$\,Hz$^{-1}$. 

We performed various tests in order to evaluate 
the accuracy of the simulations. It depends 
mainly on the mass loss rate of the central source and on the size of 
the geometrical steps adopted in the calculations. For the results presented 
in this section, the relative error on the line profile ranges from 
$\sim$ 10$^{-6}$, for mass loss rates of 10$^{-7}$ \Msold, to a few 10$^{-3}$, 
for mass loss rates of 10$^{-4}$ \Msold. 

\subsection{Freely expanding wind (scenario\,1)}
\label{freeexpansion}

\begin{figure}
\centering
\epsfig{figure=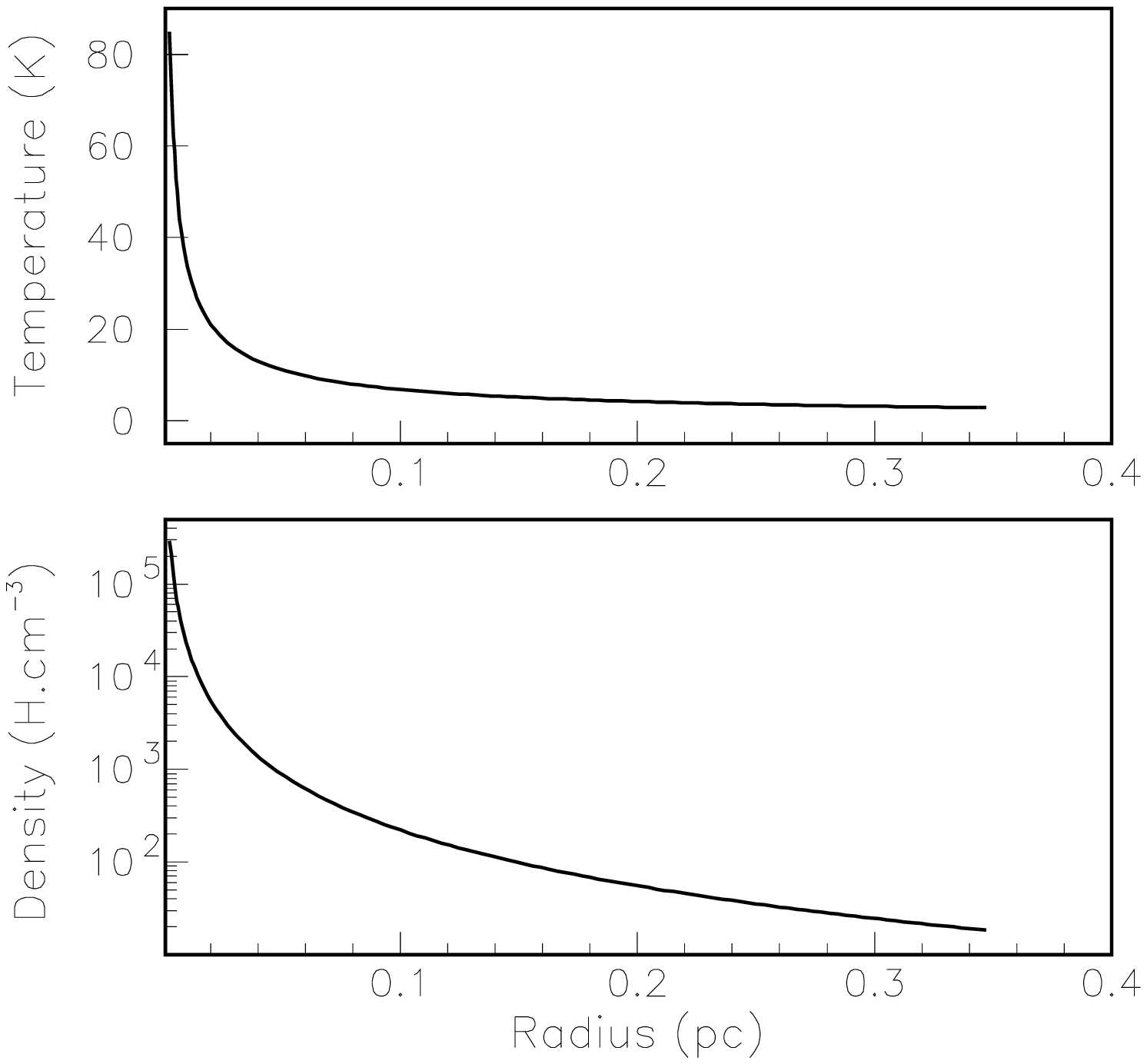,angle=0,width=8.0cm}
\caption{Density and temperature profiles for an outflow in uniform expansion 
(scenario 1, \Vexp\,=\,10\,\kms, \.M\,=\,10$^{-5}$\,\Msold).}
  \label{denstempfreewind}
\end{figure}

We consider a spherical wind in free expansion at \Vexp=10\,\kms. 
The distance is set at 200 pc, and the mass loss rate is varied from 
10$^{-7}$ to 10$^{-4}$ \Msold. We assume that the gas is composed, in number, 
of 90\% atomic hydrogen, and 10\% $^4$He. We assume a temperature dependence 
proportional to r$^{-0.7}$, r being the distance to the central star, 
out to the external boundary (0.17\,pc) where the temperature 
drops to 5\,K. This temperature of 5\,K is probably underestimated 
as the photoelectric heating by grains absorbing UV photons is expected to 
raise the temperature of the gas in the cool external layers of 
shells around stars with high mass loss rate (Sch\"oier \& Olofsson 2001). 
On the other hand, temperatures as low as 2.8\,K have been reported in some 
high mass loss rate sources (e.g. U Cam, Sahai 1990). Such low temperatures 
are only expected in the freely expanding regions of the circumstellar shells. 
The density and temperature profiles are 
illustrated in Fig.\,\ref{denstempfreewind} for the 10$^{-5}$\,\Msold ~case. 

\begin{figure}
\centering
\epsfig{figure=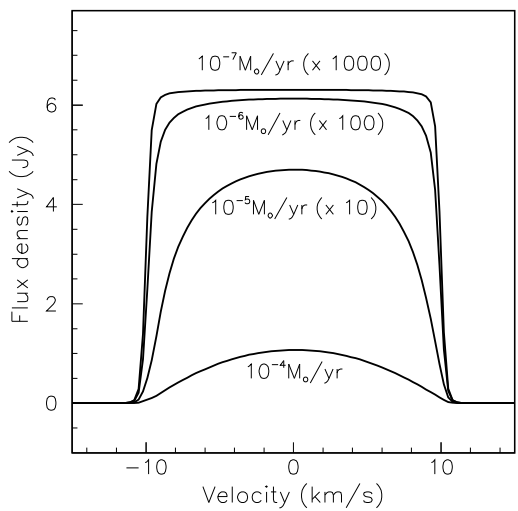,angle=0,width=8.0cm}
\caption{\HI line profiles of shells in free expansion for various mass loss 
rates with no background. The profiles for 10$^{-7}$, 10$^{-6}$, 
and 10$^{-5}$\,\Msold ~are scaled by factors 1000., 100., and 10., 
respectively. The distance is set at 200 pc.}
  \label{freenoback}
\end{figure}

In a first set of simulations (Fig.\,\ref{freenoback}) we calculate 
the integrated emission (within a diameter, $\phi=6'$) 
with no background, in order to estimate the effect of self-absorption 
(in fact the background should have a minimum brightness temperature 
of 3\,K, cf. Sect.~\ref{equations}). Self-absorption starts to play 
a clear role for 10$^{-6}$ \Msold, with an intensity that is reduced, 
and a profile that changes its shape from almost rectangular to parabolic. 
A slight asymmetry of the line profile is also present 
(although not discernible by eye in the figure), with more absorption 
on the blue side, due to the outwardly decreasing temperature, an effect which 
has already been described in the case of molecular emission from expanding 
circumstellar envelopes (Morris et al. 1985). 

\begin{figure}
\centering
\epsfig{figure=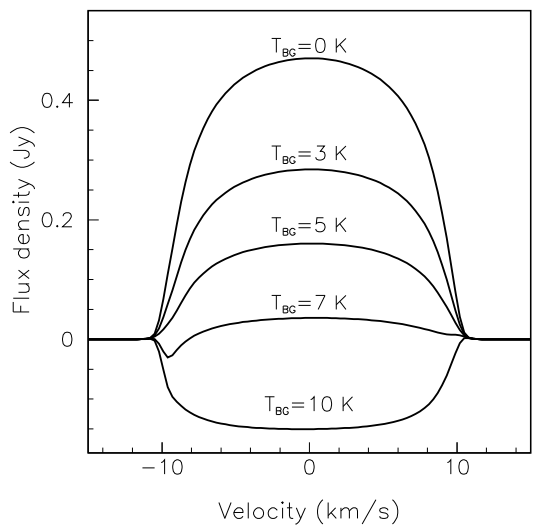,angle=0,width=8.0cm}
\caption{\HI line profiles of a shell in free expansion for 
\.M=10$^{-5}$\,\Msold, and for various background levels ($T_{BG}$= 0, 3, 5, 
7, 10\,K).}
  \label{freeback}
\end{figure}

In a second set of simulations, the mass loss rate is kept at 10$^{-5}$ 
\Msold, and the background is varied from 0\,K (as above), to 10\,K  
(Fig.\,\ref{freeback}). The effects noted previously are amplified by 
the background, in particular with an absorption developing on the blue side 
of the profile, and then extending to the complete spectral domain when 
the background temperature reaches 10\,K. 

We adopted a temperature dependence proportional to r$^{-0.7}$ which 
fits the results obtained with a radiative transfer model by Sch\"oier \& 
Olofsson (2001). A shallower dependence would increase the temperature in 
the outer layers of the circumstellar shell and thus reduce the effects of 
self-absorption, as well as the absorption of the background radiation. 
An external source of heating (e.g. by photoelectric heating) would have 
the same influence.

\subsection{Single detached shell (scenario\,2)}
\label{detachedshell}

We adopt the model developed by Libert et al. (2007). It has been shown 
to provide good spectral fits of the \HI observations obtained on sources 
with mass loss rates $\sim$\,10$^{-7}$\,\Msold ~(cf. Sect.~\ref{astroback}). 

\begin{figure}
\centering
\epsfig{figure=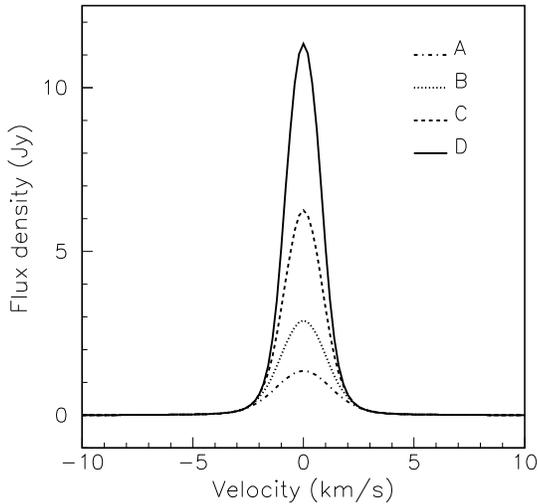,angle=0,width=8.0cm}
\caption{\HI line profiles of single detached shells for various circumstellar 
masses (A: 0.05\,\Msol, B: 0.1\,\Msol, C: 0.2\,\Msol, D: 0.4\,\Msol,), 
no background.}
  \label{detshellnoback}
\end{figure}

\begin{figure}
\centering
\epsfig{figure=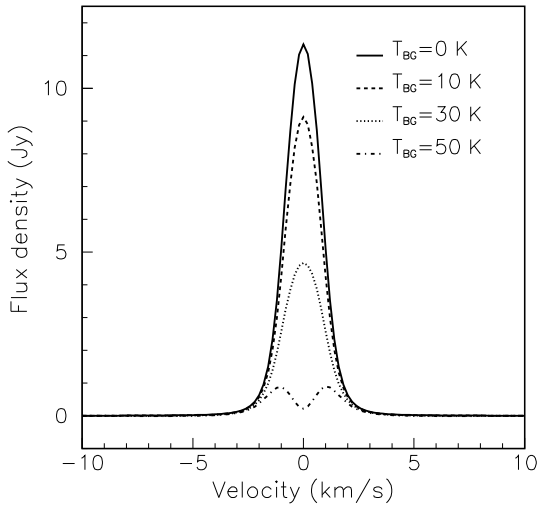,angle=0,width=8.0cm}
\caption{\HI line profiles of a single detached shell (scenario 2, case D), 
and for various background levels ($T_{BG}$=\,0\,K,\,10\,K,\,30\,K,\,50\,K).}
  \label{detshellback}
\end{figure}

\begin{figure}
\centering
\epsfig{figure=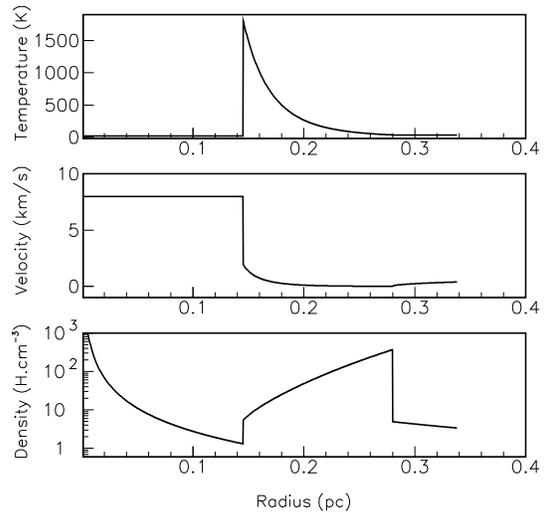,angle=0,width=8.0cm}
\caption{Density, velocity, and temperature profiles for a detached shell 
model (scenario 2, case D).}
  \label{densveltempdetshell}
\end{figure}

As in Sect.~\ref{freeexpansion}, we assume a spatially unresolved source at 
200\,pc with a mass loss rate of 10$^{-7}$ \Msold. The internal radius of 
the detached shell is set at 2.5$'$ (or 0.15\,pc). 
Similarly, we examine the dependence of the line profile for models with 
various masses in the detached shell (M$_{DT,CS}$) and various background 
levels (Figs.~\ref{detshellnoback} and \ref{detshellback}). 
The parameters of the four cases illustrated in Fig.~\ref{detshellnoback} 
are given in Table~\ref{case2parameters}. The free-wind expansion velocity 
is taken to be \Vexp =\,8\,\kms. At the termination shock the downstream 
temperature is given by T$_f \sim$ (3 $\mu\,m_H$)/(16 k) \Vexp$^2 \sim$ 1800 K 
(equ. 6.58 in Dyson \& Williams 1997) with $m_H$, the mass of the hydrogen 
atom, and $\mu$, the mean molecular weight. For the temperature profile 
inside the detached shell we use the expression (9) in Libert et al. (2007) 
with a temperature index, $a=-6.0$. 
The temperature is thus decreasing from  $\sim$\,1800\,K, to T$_f$, at the 
interface with ISM, r$_f$. The density, velocity and temperature profiles are 
illustrated in Fig.\,\ref{densveltempdetshell} for case D. 

Self-absorption within the detached shell has a limited effect, with a 
reduction ranging from 1\% (model A) to 20\% (model D), as compared to the 
optically thin approximation (Fig.~\ref{detshellnoback}). 
However, taking into account the background 
introduces a much larger effect (Fig.~\ref{detshellback}).

The results depend on the adopted parameters in the model (mainly 
internal radius, expansion velocity and age). Smaller internal radius and 
expansion velocity, and/or longer age would lower the average temperature in 
the detached shell. This would increase the effect of self-absorption, as 
well as that of the background absorption. The line profiles simulated 
with the A and B-cases represented in Fig.~\ref{detshellnoback} provide 
a good approximation to several observed \HI line profiles 
(Libert et al. 2007, 2010, Matthews et al. 2013).

\begin{table}
\centering
\caption{Model parameters (scenario 2), d = 200 pc, \Vexp = 8\,\kms, and 
\.M\,=\,10$^{-7}$ \Msold.}
\begin{tabular}{ccccc}
\hline
case  & age (yr) & r$_f$(arcmin)  & T$_f$ (K) & M$_{DT,CS}$ (\Msol)\\
\hline
A     & 5$\times$10$^{5}$  & 3.85  & 135 &  0.05 \\
B     &  10$^{6}$          & 4.14  &  87 &  0.1  \\
C     & 2$\times$10$^{6}$  & 4.47  &  55 &  0.2  \\
D     & 4$\times$10$^{6}$  & 4.83  &  35 &  0.4  \\
\hline
\end{tabular}
\label{case2parameters}
\end{table}

\begin{figure}
\centering
\epsfig{figure=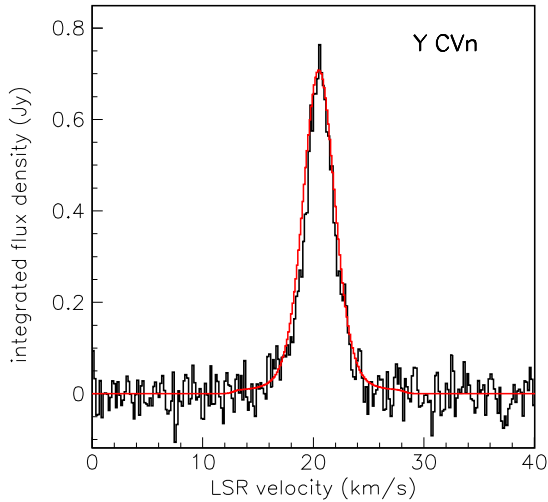,angle=0,width=8.0cm}
\caption{Y CVn integrated spectrum (Libert et al. 2007) and fit obtained 
by Hoai (2015) with scenario\,2 (d\,=\,321\,pc, \.M\,=\,1.3$\times$10$^{-7}$ 
\Msold, age\,=\,7$\times$10$^{5}$ years).}
  \label{YCVn}
\end{figure}

As an illustration, we reproduce on Fig.~\ref{YCVn} the spatially integrated 
profile of Y CVn observed by Libert et al. (2007) together with a recent fit 
obtained by Hoai (2015). For this fit, a distance of 321 pc (van Leeuwen 
2007), a mass loss rate of 1.3$\times$10$^{-7}$ \Msold, and a duration of 
7$\times$10$^{5}$ years have been adopted. These parameters differ from those 
adopted by Matthews et al. (2013), who assumed 1.7$\times$10$^{-7}$ \Msold 
~and a distance of 272 pc (Knapp et al. 2003). 
However, by adopting a lower mass loss rate, and conversely, 
a longer duration, Hoai (2015) can fit the spatially resolved 
spectra obtained by the VLA and solve the problem faced by Matthews et al.  
at small radii. A difference between the mass loss rate estimated from 
CO observations and that adopted in the model may have several reasons, 
for instance an inadequate CO/H abundance ratio.

\subsection{Villaver et al. model (scenario\,3)}
\label{villaver}

Villaver et al. (2002) have modeled the dynamical evolution of circumstellar 
shells around AGB stars. The temporal variations of the stellar winds are 
taken from the stellar evolutionary models of Vassiliadis \& Wood (1993). 

\begin{figure*}
\centering
\epsfig{figure=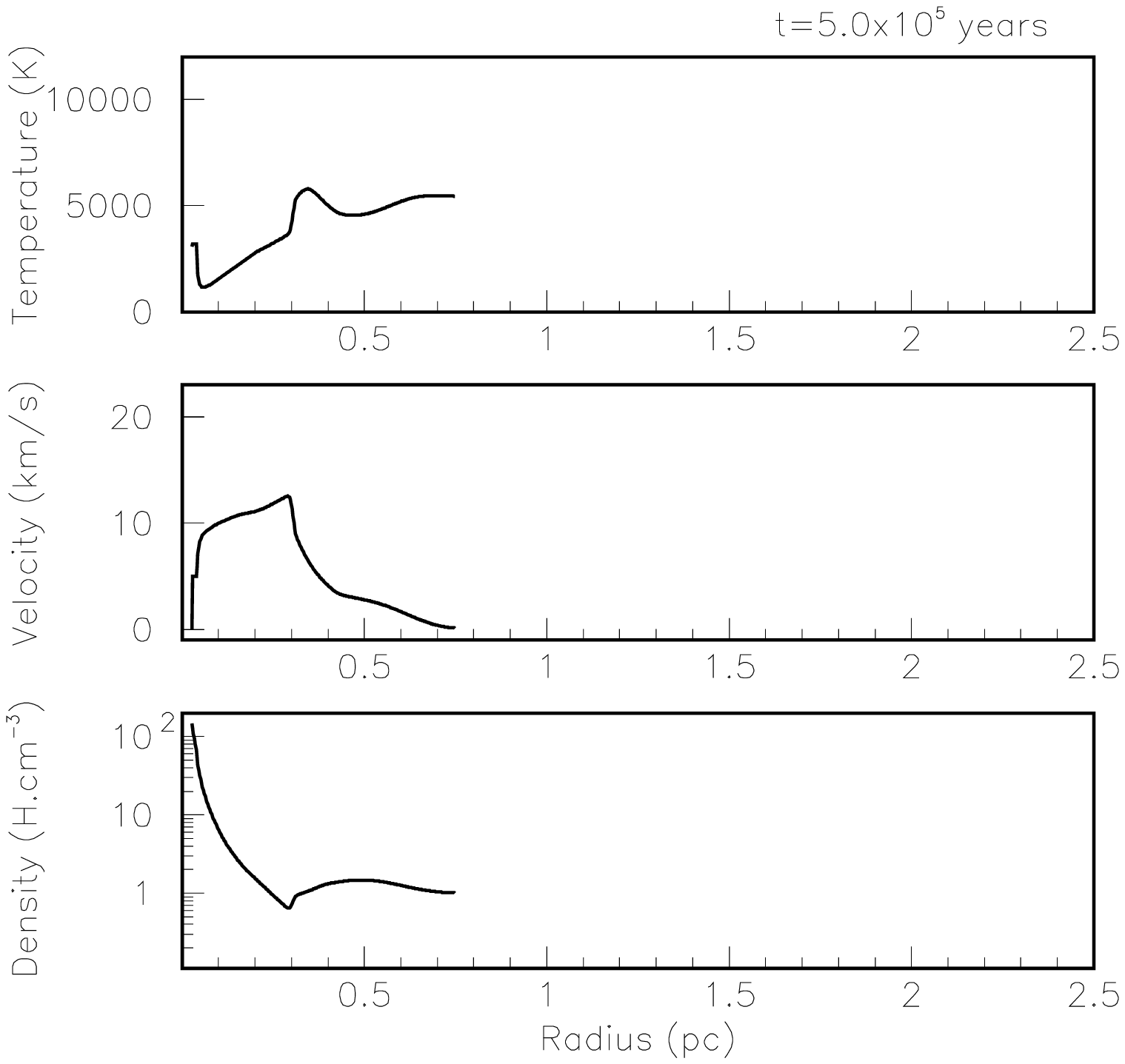,angle=0,width=5.8cm}
\epsfig{figure=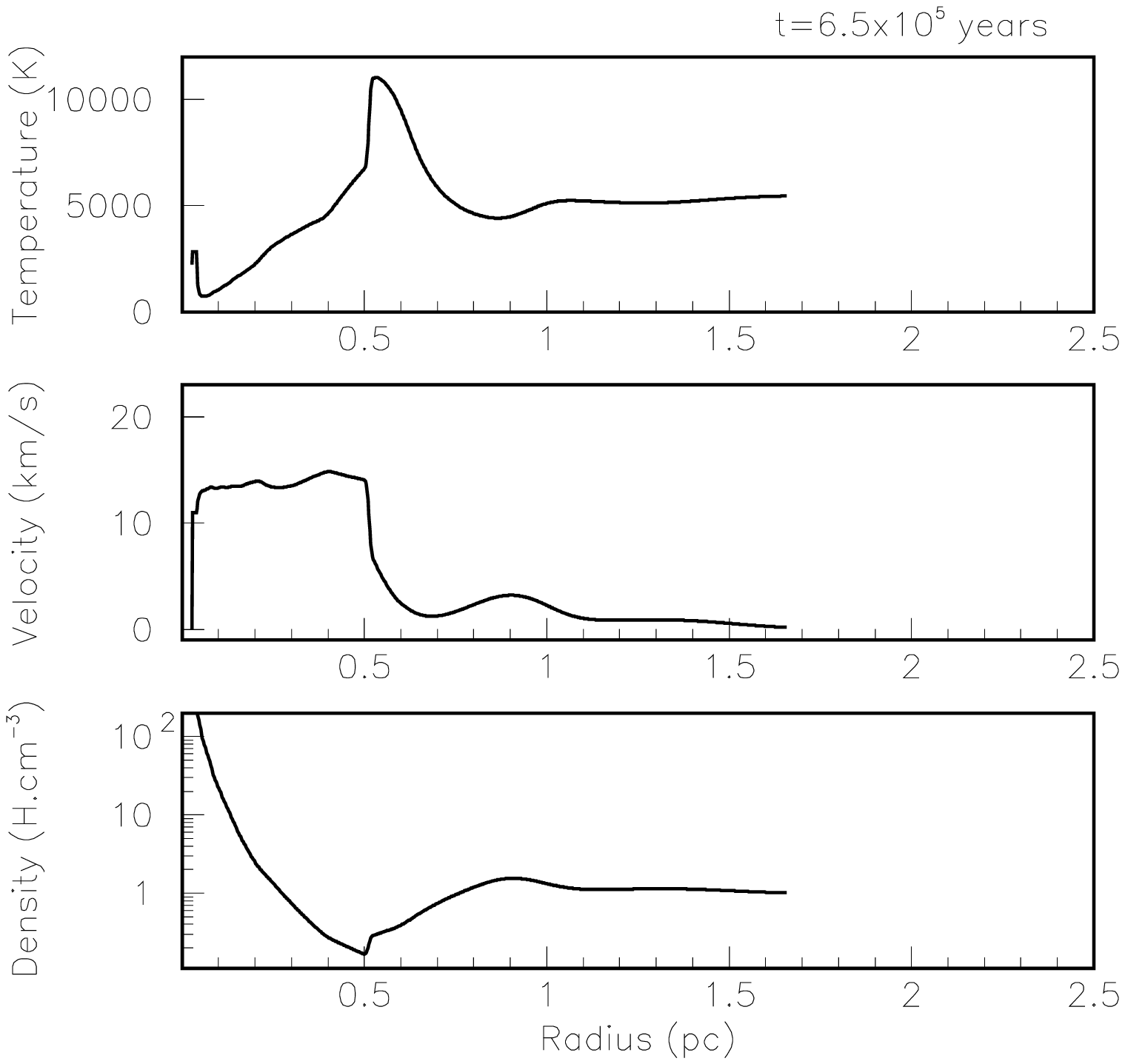,angle=0,width=5.8cm}
\epsfig{figure=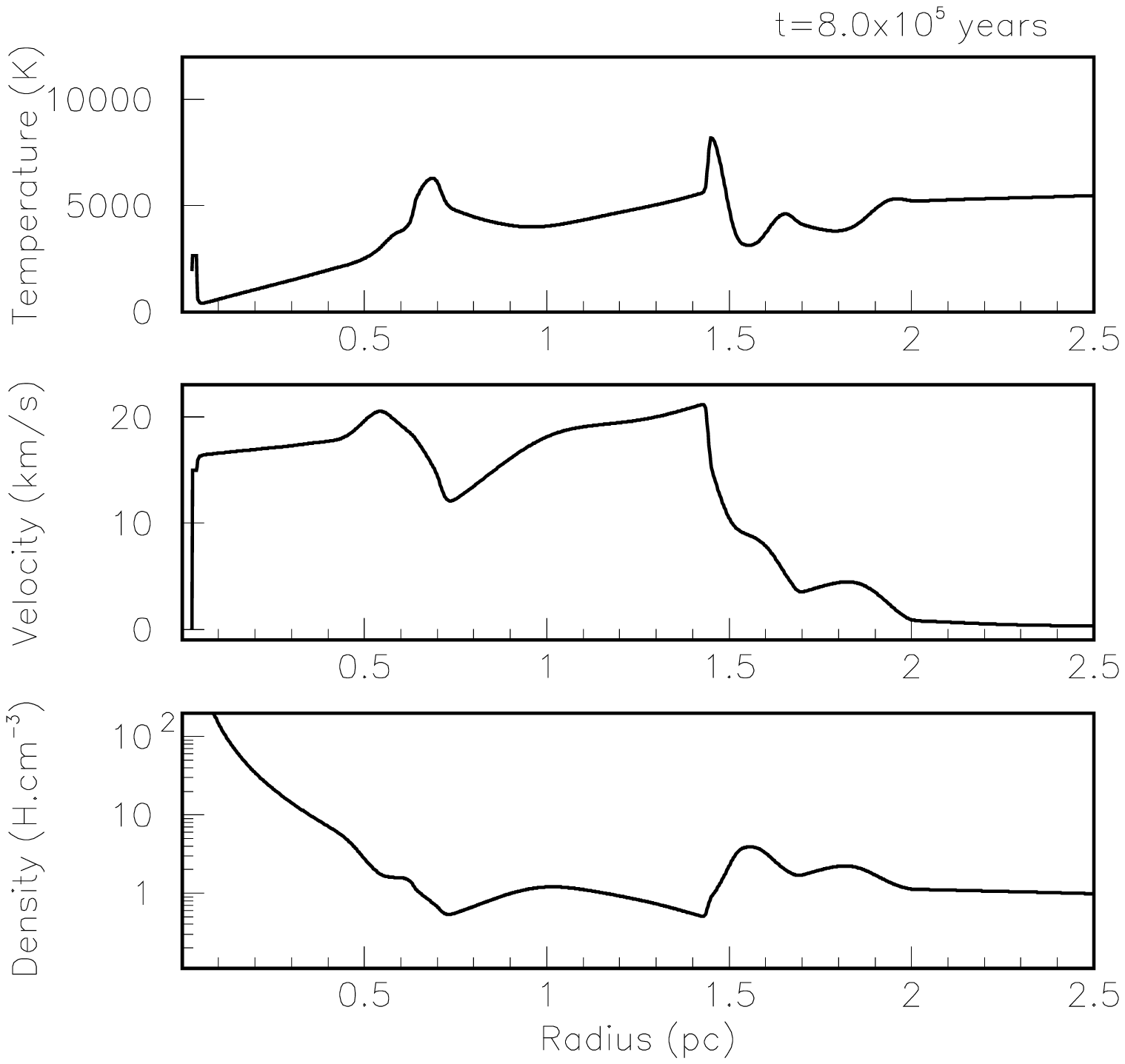,angle=0,width=5.8cm}
\caption{Density, velocity, and temperature profiles for the Villaver et 
al. (2002) model (scenario 3) at three different epochs (5.0 $\times$10$^5$
years (left), 6.5 $\times$10$^5$ years (center), 8.0 $\times$10$^5$ years 
(right).}
  \label{densveltempVillaver}
\end{figure*}

For our \HI simulations, we used the 1.5-\Msol ~circumstellar shell models of 
Villaver et al. (2002) at various times of the thermally pulsing AGB 
(TP-AGB) evolution. We selected 
the epochs at 5.0, 6.5, and 8.0$\times$10$^5$ years, which correspond 
to the first two thermal pulses, and then to the end of the fifth (and last) 
thermal pulse. The density, velocity and temperature profiles are 
illustrated in Fig.\,\ref{densveltempVillaver}.
For these models, which can reach a large size (with radii 
of 0.75, 1.66 and 2.5 pc, respectively), we adopt a distance of 1 kpc 
(implying a diameter of up to 17$'$).

\begin{figure}
\centering
\epsfig{figure=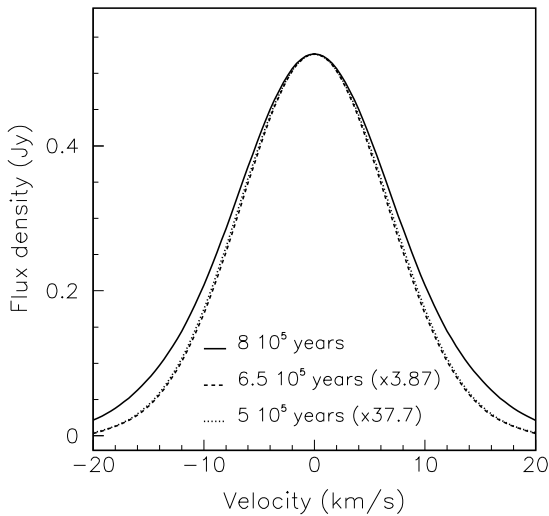,angle=0,width=8.0cm}
\caption{\HI line profiles of a circumstellar shell model around a 1.5-\Msol 
~star during the evolution on the TP-AGB phase (5.0, 6.5, 8.0$\times$10$^5$ 
years, Villaver et al. 2002), no background. The distance is set at 1000 pc. 
The first two profiles have been scaled by 37.7 and 3.87, respectively, in 
order to help the comparison between the different line profiles.}
  \label{villavernoback}
\end{figure}

The results are shown in Fig.~\ref{villavernoback}. In this scenario, 
the temperature in the circumstellar environment is maintained at high 
values due to the interactions between the successive shells 
(Fig.~\ref{densveltempVillaver}). 
The shape of the line profile is thus dominated by thermal broadening, 
and does not depend much on the epoch which is considered 
(although the intensity of the emission depends strongly on time, 
together with the quantity of matter expelled by the star). 

For the same reason, these results do not depend much on the background 
($<$5\% for $T_{BG}$=100\,K). Indeed in the models the temperature of the gas 
in the circumstellar shell always stays at a high level ($> 10^3$ K, 
except close to the central star in the freely expanding region).

The predictions obtained with this scenario, in which wind-wind interactions 
are taken into account, differ clearly from those obtained with the previous 
scenario, in which the detached shell is assumed to result from a 
long-duration stationary process, by a much larger width of the line profiles 
(full-width at half-maximum, FWHM\,$\sim$\,16\,\kms). This large width 
in the simulations for scenario n$^{\circ}$3 results mainly from the thermal 
broadening, and also, but to a lesser extent, from the kinematic broadening 
(cf. Fig.~\ref{densveltempVillaver}).   

\section[]{Discussion}
\label{discussion}

\subsection{Optically thin approximation}
\label{thinapprox}

If absorption can be neglected, the intensity becomes proportional 
to the column density of hydrogen. For a source at a distance $d$, 
the mass in atomic hydrogen ($M_{\rm HI}$) can be derived from the integrated 
flux density through the standard relation (e.g. Knapp \& Bowers 1983): \\ 
$M_{\rm HI} = 2.36\times10^{-7}d^{2} \int S_{\rm HI} dV$, \\
in which $d$ is expressed in pc, $V$ in \kms, $S_{\rm HI}$ in Jy, 
and $M_{\rm HI}$ in solar masses (\Msol).

\begin{figure}
\centering
\epsfig{figure=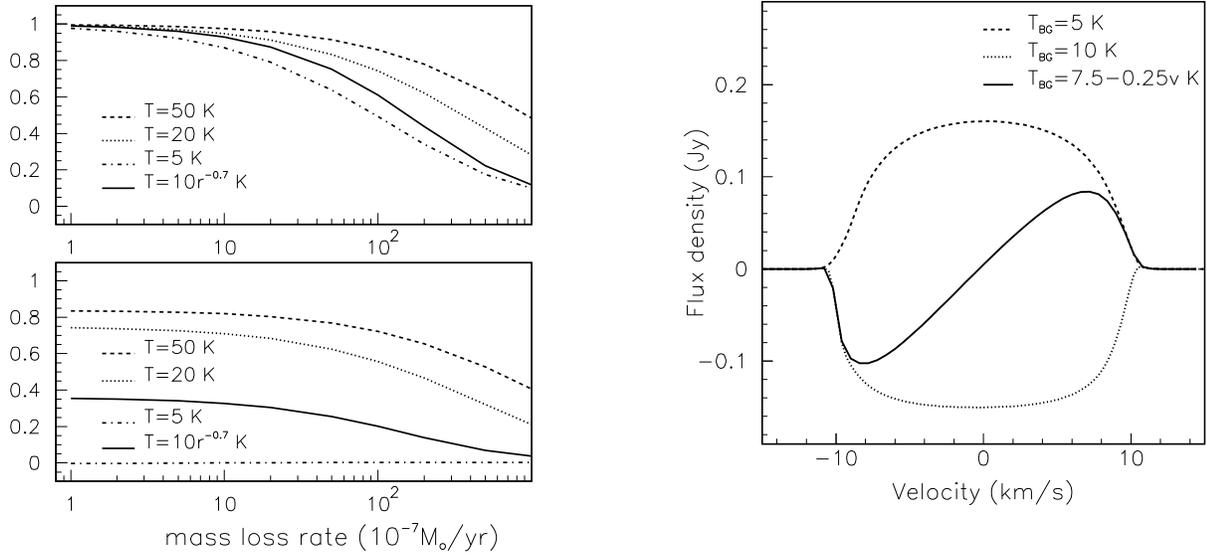,angle=0,width=8.6cm}
\caption{Ratio between the ``estimated'' mass in atomic hydrogen and the real 
mass for the freely expanding wind case (scenario n$^{\circ}$\,1) 
with mass loss rates ranging from 10$^{-7}$ to 10$^{-4}$ \Msold 
~(see Sect.~\ref{thinapprox}) and different cases of temperature 
dependence (see text). 
Upper panel: no background. Lower panel: with a 5-K background.}
  \label{hydrogenestimate}
\end{figure}

Our calculations allow us to estimate the error in the derived \HI mass 
of circumstellar envelopes that is incurred from the assumption that the 
emission is optically thin and not affected by the background. As an example, 
in Fig.~\ref{hydrogenestimate}, we show the ratio between the estimated mass 
(using the standard relation) and the exact mass in atomic hydrogen. 
The case without background illustrates the effect of self-absorption 
within the circumstellar shell for different mass loss rates. 
We adopt a freely expanding wind with a temperature profile in r$^{-0.7}$ 
(as in Sect.~\ref{freeexpansion}), with r expressed in arcmin.,  
or a constant temperature (5\,K, 10\,K, 20\,K). 
The ratio clearly decreases with decreasing temperature in the circumstellar 
shell, increasing mass loss rate and increasing background temperature. 

In the constant temperature case with T\,=\,5\,K and $T_{BG}$=\,5\,K, 
the line profiles should be flat (cf. radiative transfer equation in 
Sect.~\ref{equations}), and thus the ``estimated'' masses, exactly null. 
Our numerical calculations agree with this prediction to better 
than 3$\times10^{-3}$, for mass loss rates up to 10$^{-4}$\,\Msold. 

The standard relation used for estimating the mass in atomic hydrogen should 
obviously be handled with caution in the case of the freely expanding wind 
scenario (n$^{\circ}$\,1). On the other hand, our calculations show that 
the deviation is much smaller for the two other scenarios (and basically 
negligible for scenario n$^{\circ}$\,3). This is mainly an effect of the high 
temperature in the detached shells resulting from the wind-wind interactions. 

\subsection{Spectral variations of the background}
\label{specvarbg}
 
\begin{figure}
\centering
\epsfig{figure=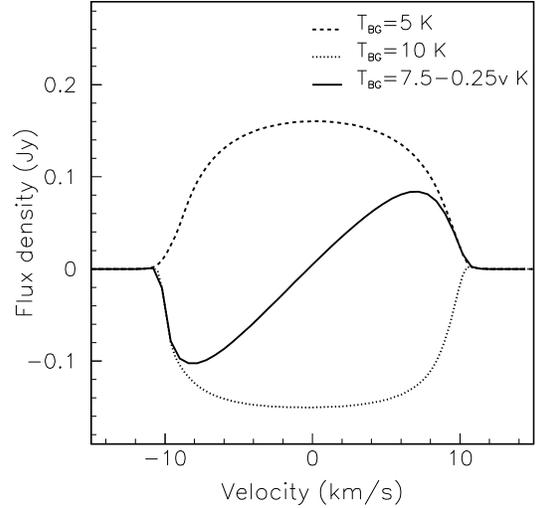,angle=0,width=8.0cm}
\caption{Effect of a background intensity varying linearly from 10\,K to 5\,K 
across the line profile for a scenario\,1 model with \.M=10$^{-5}$\,\Msold. 
The curves labelled '$T_{BG}$=\,5\,K', and '$T_{BG}$=\,10\,K', 
are reproduced from Fig.~\ref{freeback}.}
  \label{variablebackground}
\end{figure}

The \HI absorption produced by cold galactic gas in the foreground of bright 
background emission may be shifted towards the velocity with highest 
background (cf. Levinson \& Brown 1980). 
To illustrate this effect in the case of 
circumstellar shells, in Fig.~\ref{variablebackground}, we show the results 
of our simulations for a 10$^{-5}$\,\Msold ~freely expanding wind, as in 
Sect.~\ref{freeexpansion}, and a background temperature varying linearly 
between 10\,K at $-$10~\kms, and 5\,K at +10~\kms. The absorption is clearly 
shifted towards velocities with highest background. One notes also 
that the emission is shifted towards velocities with lowest background. 

In case of a intense and spectrally structured background, 
some care should be exercised when comparing the \HI line centroids 
with the velocities determined from other lines. 

\subsection{Comparison with observations}
\label{comp}
 
Freely expanding winds have been definitively 
detected in the \HI line in only two red 
giants: Y\,CVn (Le Bertre \& G\'erard 2004) and Betelgeuse (Bowers \& Knapp 
1987). The corresponding emission is relatively weak and difficult to detect. 
Data obtained at high spatial resolution reveal a double-horn profile 
(e.g. Bowers \& Knapp 1987). It is worth noting that a high velocity expanding 
wind (\Vexp\,$\sim$\,35\,\kms) has also been detected around the classical 
Cepheid $\delta$\,Cep (Matthews et al. 2012).
A pedestal is suspected in a few \HI line profiles that could be due the 
freely expanding region (G\'erard \& Le Bertre 2006, Matthews et al. 2013). 
The first scenario might also be interesting for interpreting 
sources in their early phase of mass loss, or for sources, 
at large distance from the Galactic Plane, embedded in a low pressure ISM. 

In general, sources which, up to now, 
have been detected in \HI show quasi-gaussian line profiles  
of FWHM\,$\sim$\,2-5\,\kms ~(G\'erard \& Le\,Bertre 2006, Matthews et al. 
2013), a property which reveals the presence of slowed-down detached shells. 
These profiles are well reproduced by simulations based on the scenario 
n$^{\circ}$2 presented in Sect.\,\ref{detachedshell}, assuming mass loss rates 
of a few 10$^{-7}$ \Msold, and durations of a few 10$^{5}$ years. 
In particular, for Y CVn and Betelgeuse, the main \HI component has 
a narrow line profile ($\sim$\,3\,\kms) and is well reproduced by this kind 
of simulation (Libert et al. 2007, Le\,Bertre et al. 2012). 

Sources with large mass loss rates ($\geq 5\times 10^{-7}$\,\Msold)
have rarely been detected (with the notable exceptions of 
IRC\,+10216 and AFGL\,3068, see below). 
The simulations presented in Sect.~\ref{villaver} show the line profiles that 
sources, such as those predicted by Villaver et al. (2002), should exhibit 
at the end of the thermal-pulse phase, with large mass loss rates, 
and with interaction with the local ISM. 
In these models, in which the evolution of the central star is integrated, 
the circumstellar envelopes result from several interacting shells, 
as well as from the ISM matter which has been swept-up. Shocks between 
successive shells maintain a high gas temperature ($\sim$\,4000\,K). 

For these models the calculated line profiles are not seriously affected by 
the background level, and the flux densities are large enough for allowing 
a detection up to a few kpc. For instance, in the GALFA-\HI survey 
(Peek et al. 2011), the 3-$\sigma$ detection limit for a point source in 
a 1\,\kms ~channel is $\sim$\,30\,mJy. Saul et al. (2012) have detected 
many compact isolated sources in this survey. However, at the present stage, 
none could be associated with an evolved star (Begum et al. 2010). 

Furthermore, several sources with high mass loss rates, such 
as IRC\,+10011 (WX Psc), IK\,Tau (NML\,Tau) or AFGL\,3099 (IZ Peg) 
which are observed at high galactic latitude, 
with an expected low interstellar \HI background, remain undetected 
(G\'erard \& Le\,Bertre 2006, Matthews et al. 2013). The simulations that 
we have performed based on the three different scenarios considered in 
this work cannot account for such a result. It seems therefore that, 
in sources with large mass loss rates ($\geq 5\times 10^{-7}$\,\Msold), 
hydrogen is generally not in atomic, but rather in molecular form. 

Glassgold \& Huggins (1983) have discussed the H/H$_2$ ratio in the 
atmospheres of red giants. They find that for stars with photospheric 
temperature \Tstar $\geq$ 2500 K, most of the hydrogen should be in atomic 
form, and the reverse for \Tstar $\leq$ 2500 K. Winters et al. (2000) 
find that there is an anti-correlation between \Tstar ~and the mass loss 
experienced by long period variables. It seems likely that stars 
having a mass loss rate larger than a few 10$^{-7}$\,\Msold ~have 
also generally a low photospheric temperature, with \Tstar\,$\leq$\,2500\,K, 
and thus a wind in which hydrogen is mostly molecular. 

Recently, Matthews et al. (2015) have reported the detection of atomic 
hydrogen in the circumstellar environment of IRC\,+10216, a prototype of 
a mass losing AGB star at the end of its evolution with \.M$\sim$ 
2$\times$10$^{-5}$ \Msold. The observed morphology, 
with a complete ring of emission, is in agreement with the predictions 
of Villaver et al. (2002, 2012). They find that atomic hydrogen represents 
only a small fraction of the expected mass of the circumstellar 
environment ($<$\,1\,\%), supporting a composition dominated by molecular 
hydrogen. Unfortunately, a reliable line profile could not be extracted due 
to the low level of the emission and to a patchy background. The detection 
of \HI over a spectral range $\sim$\,10\,\kms ~suggests a line width larger 
than commonly observed in evolved stars, which would make it compatible 
with scenario n$^{\circ}$3. G\'erard \& Le\,Bertre (2006) have reported 
the possible detection of AFGL\,3068, another carbon star with high mass 
loss rate ($\sim$\,10$^{-4}$\,\Msold). In this case, also, the line width  
($\sim$\,30\,\kms) is larger than expected for scenario n$^{\circ}$2, and 
might be better explained by scenario n$^{\circ}$3. Another possibility 
for this source which is at a large distance from the Galactic Plane 
(z\,$\sim$\,740\,pc) would be that we are mostly detecting a freely expanding  
wind not slowed down by its local ISM (i.e. scenario n$^{\circ}$1).

If the atomic hydrogen is of atmospheric origin (a fraction of 1\,\% 
is expected for a star with an effective temperature of 2200\,K, Glassgold 
\& Huggins 1983), its abundance should correspondingly be scaled down 
in our radiative transfer simulations. The effect of the optical depth 
on the line profiles could be considerably reduced for such a case. 
For stars with lower effective temperature (\Tstar $\leq$\,2200\,K), atomic 
hydrogen might also be present in the external regions of circumstellar 
envelopes as a result of the photodissociation of molecular hydrogen by 
UV photons from the interstellar radiation field (Morris \& Jura 1983). 

\subsection{Case of a resolved source}
 \label{resolvedsource}

\begin{figure}
\centering
\epsfig{figure=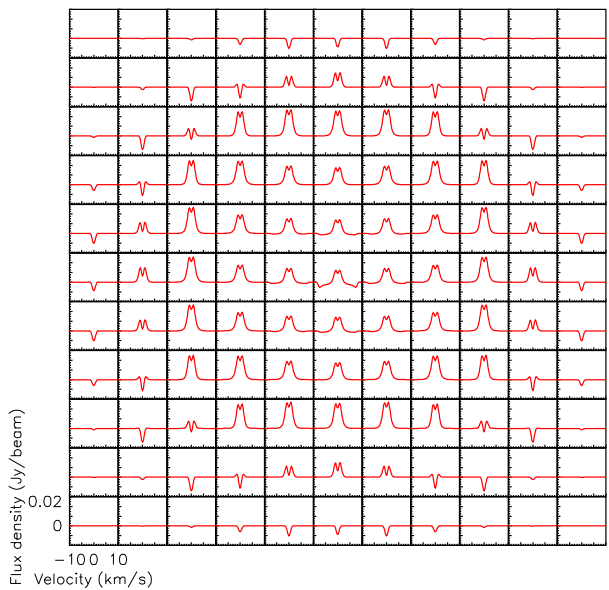,angle=0,width=9.2cm}
\caption{\HI spectral map for a detached shell (case D in 
Table~\ref{case2parameters} with $T_{BG}$=\,50\,K), assuming a gaussian 
beam of FWHM\,=\,1$'$. Steps are 1$'$ in both directions.}
  \label{spectralmapdetshell}
\end{figure}

We have concentrated our study on the prediction of spatially integrated 
spectra. However, circumstellar envelopes may reach a large size 
($\sim$\,2-3\,pc, Villaver et al. 2002), and thus have a large extent over 
the sky. Also, interferometers provide a larger spatial resolution 
than single-dish antennas. It is thus interesting to examine how the line 
profile may vary as a function of position. As an example, in 
Fig.~\ref{spectralmapdetshell}, we show a spectral map that would be obtained 
for a detached shell observed over a background with $T_{BG}$=\,50\,K 
(scenario\,2). The line appears mostly in emission and, as expected, 
delineates the detached shell. However, in the case of a high background 
level, the line appears also in absorption, in particular in the external 
part of the detached shell where the lines of sight cross regions with gas 
at low temperature. 

Spatially resolved \HI studies, with a careful subtraction of the background 
emission, may thus reveal spectral signatures that hold information on 
their physical conditions. 
Such signatures could help to constrain the physical properties of 
the gas in a region where molecules are absent or not detectable.

\section[]{Prospects}
\label{prospects}
 
We have simulated \HI 21-cm line profiles for mass losing AGB stars 
expected for different scenarios assuming spherical symmetry.

However, AGB sources are moving through the ISM and their shells may be 
partially stripped by ram pressure (Villaver et al. 2003, 2012).  
As a consequence of the interaction a bow-shock shape appears in the direction 
of the movement, but also a cometary tail is formed which is fed directly from 
the stellar wind and from material stripped away from the bow shock. The 
cooling function and the temperature assumed for the wind have an important 
effect on the formation of the tail as shown in Villaver et al. (2012).
Higher density regions formed behind the star will cool more
efficiently and will collapse against the ISM pressure, allowing
the formation of narrow tails. 

G\'erard \& Le Bertre (2006) have reported shifts of the \HI emission 
in velocity as well as in position for several sources. 
Matthews et al. (2008) have reported a shift in velocity for different 
positions along the tail of Mira (see also X Her, Matthews et al. 2011). These 
effects can also affect the \HI line profiles, and thus the detectability.
In addition, material lost by the AGB star should be spread along a tail 
that may reach a length of 4 pc, as in the exceptional case of Mira. 
On the other hand, Villaver et al. (2012) show that for sources with 
large mass loss rates at the end of their evolution, dense shells could still 
be found close to the present star position. 

We have assumed a background with a constant brightness. Of course, 
as explained in Sect.~\ref{equations}, this applies only to the cosmic 
background, and to a lesser extent to the galactic continuum emission. 
It does not apply to the galactic \HI emission which may show spatial 
structures of various kinds. The resulting effect may be more complex than 
that simulated in Sect.~\ref{simul}. For instance, an absorption line could 
form preferentially at the position of a peak of galactic \HI emission 
(a radiative transfer effect). Such a phenomenon may affect the predictions 
presented in 
Sect.~\ref{resolvedsource}. Therefore a good description of the background 
will also be needed to model the observed line profiles. Such input may be 
obtained through frequency-switched observations for the galactic \HI 
component, and through the surveys of the continuum at 21 cm which are 
already available (see Sect.~\ref{equations}). 

It has to be noted that, in the position-switched mode of observation, 
the intrinsic line profile of the stellar source can also be spoiled by the 
patchiness of the galactic background emission (observational artifact). 
The main source of confusion is the galactic \HI emission which is structured 
spatially and spectrally. The classical position-switched mode of observation 
is not always efficient to correct the 21-cm spectra from the emission 
that is not directly associated with the star. More sophisticated methods 
with 2D mapping might be needed for subtracting the contaminating emission 
in these cases. Interferometric observations have the advantage of filtering 
the large-scale galactic emission. However, one should care that an intrinsic 
circumstellar emission is not also subtracted in this mode of observation. 
Also some artifacts may arise from incomplete spatial sampling of the large 
scale emission, as illustrated by the case of TX Psc (Matthews et al. 2013). 
If feasible, an excellent u-v coverage combined with maps from a single dish 
telescope providing small spacings has to be obtained. Also, 
for circumstellar shells angularly larger than the primary beam 
of the interferometer, mosaicked observations combined with single dish maps 
are needed.
 
Another caveat is that, when the distance to the source becomes larger,  
the foreground ISM material may play the role of an absorbing layer  
of growing importance (Zuckerman et al. 1980). The circumstellar shell 
line profile may thus be distorted by absorption due to the foreground 
cold material that shares the same radial velocity range. 

Sources with high mass loss rate ($\sim$\,10$^{-6}$--10$^{-5}$\,\Msold) tend 
to be concentrated towards the galactic plane. They are expected to dominate 
the contribution of AGB stars to the replenishment of the ISM (Le\,Bertre et 
al. 2003). The recent detection of IRC\,+10216 by Matthews et al. (2015) 
shows that atomic hydrogen should be present in these sources and that 
the \HI line at 21 cm can be used to probe the morphology and the kinematics 
of stellar matter decelerated at large distance from the central star.
However, as discussed above, when the background is large, a proper modelling 
of the line profiles will be necessary.

\section[]{Summary and Conclusions}
\label{conclusion}

We have simulated \HI 21-cm line profiles expected for several different 
scenarios representing different evolutionary stages of evolved stars, and 
thus corresponding to different AGB circumstellar structures. We have relaxed 
the optically thin hypothesis which was assumed in previous works, 
and included the emission from the background. 

Self-absorption may be important in freely expanding circumstellar shells, 
as well as in some detached shells resulting from the interaction of the 
stellar winds with the local ISM. The \HI line profile may also be affected by 
the background level and by the spectral profile of this background emission. 

The numerical simulations that we have performed show that, under certain 
conditions, the observed \HI 21-cm flux densities from mass-losing stars 
can be significantly reduced by taking into account optical depth effects 
and the presence of the background emission, but not to such a level 
such as to account for the non-detection of several sources. 
Therefore, one should consider that molecular hydrogen instead 
of atomic hydrogen likely dominates in  sources with 
high mass loss rates ($\geq$ few 10$^{-7}$\,\Msold), probably an effect of 
their low atmospheric temperature. Still, the recent results of Matthews et 
al. (2015) show that the \HI line at 21 cm can be a useful probe of the outer 
regions of sources with low stellar effective temperature ($<$\,2500\,K). 

For sources with mass loss rates $\sim$\,10$^{-7}$\,\Msold, which are detected 
in H\,{\sc {i}}, the global agreement between the observed line profiles and 
the simulations based on the second scenario suggests that their central stars 
undergo mass loss smoothly over several 10$^5$\,years.

\section*{Acknowledgments}
We thank Pierre Darriulat and Jan Martin Winters for their continuous support 
and kind encouragements. 
We are also grateful to N. Cox and A. J. van Marle, the organisers 
of the Lorentz workshop on Astrospheres (Leiden, 9-13 dec. 2013), where 
the ideas developed in this paper started to take shape. DTH and PTN thank 
the French Embassy in Hanoi and the CNRS/IN2P3 for financial support. 
Financial and/or material support from the Institute for Nuclear Science and 
Technology, Vietnam National Foundation for Science and Technology Development 
(NAFOSTED) under grant number 103.08-2012.34 and World Laboratory 
is gratefully acknowledged. LDM is supported by grant AST-1310930 from the 
National Science Foundation. EV acknowledges Spanish Ministerio de Econom\'ia 
y Competitividad funding under grant AYA2013-45347P. TL acknowledges 
financial support by the CNRS programmes ASA and PCMI.

\label{lastpage}

\end{document}